%% file: eydle-paper-IEEE.tex
\documentclass[conference]{IEEEtran}
\IEEEoverridecommandlockouts
\usepackage{cite}
\usepackage{amsmath,amssymb,amsfonts}
\usepackage{algorithmic}
\usepackage{graphicx}
\usepackage{textcomp}
\usepackage{xcolor}
\usepackage[hidelinks]{hyperref}
\usepackage{breakurl}
\usepackage{caption}
\usepackage{subcaption}
\usepackage[draft,layout={index,margin}]{fixme}
\usepackage{booktabs}
\usepackage{flushend}

\def\BibTeX{{\rm B\kern-.05em{\sc i\kern-.025em b}\kern-.08em
    T\kern-.1667em\lower.7ex\hbox{E}\kern-.125emX}}
\begin{document}


\title{Distributed Deep Learning Using  Volunteer Computing-Like Paradigm}

\author{
\IEEEauthorblockN{Medha Atre}
\IEEEauthorblockA{\textit{Eydle Inc.} \\
medha.atre@eydle.com}
\and
\IEEEauthorblockN{Birendra Jha}
\IEEEauthorblockA{\textit{Eydle Inc.} \\
birendra.jha@eydle.com}
\and
\IEEEauthorblockN{Ashwini Rao}
\IEEEauthorblockA{\textit{Eydle Inc.} \\
ashwini.rao@eydle.com}
}

\maketitle

\begin{abstract}
Use of Deep Learning (DL) in commercial applications such as image classification, sentiment analysis and speech recognition is increasing. When training DL models with large number of parameters and/or large datasets, cost and speed of training can become prohibitive. Distributed DL training solutions that split a training job into subtasks and execute them over multiple nodes can decrease training time. However, the cost of current solutions, built predominantly for cluster computing systems, can still be an issue. In contrast to cluster computing systems, Volunteer Computing (VC) systems can lower the cost of computing, but applications running on VC systems have to handle fault tolerance, variable network latency and heterogeneity of compute nodes, and the current solutions are not designed to do so. We design a distributed solution that can run DL training on a VC system by using a data parallel approach. We implement a novel asynchronous SGD scheme called VC-ASGD suited for VC systems. In contrast to traditional VC systems that lower cost by using untrustworthy volunteer devices, we lower cost by leveraging preemptible computing instances on commercial cloud platforms. By using preemptible instances that require applications to be fault tolerant, we lower cost by 70-90\% and improve data security. 
\end{abstract}


\input{intro}

\input{background}

\input{sysdesign}

\input{expt}

\section{Limitations}\label{sec:limitations}
We conducted experiments on CPU instances. In deep learning, training with GPUs is popular, but GPUs are much more expensive than CPU instances. Preemptible GPU computing instances are also available at 70-90\% discount. We believe we can apply our design to GPU instances as well. Both BOINC and Tensorflow support computing with GPUs. For instances with a single GPU, as long as the required GPU drivers are installed, training can run without any changes because the client-side Tensorflow training code will use the GPU by default. For instances with multiple GPUs, we may need minor modifications to the client-side code. We also need to consider challenges such as data sharing among GPUs. 

We validated our design using CIFAR10 image classification benchmark problem. We are conducting experiments using larger problems such as ImageNet. We plan to run experiments for other deep learning problems such as NLP, machine translation and time-series forecasting because they can impose new challenges. For example, the size of the training data for image classification is usually large and has to be managed using compression and caching. However, the size of training data for time-series forecasting is often small. Furthermore, image classification problems are more amenable to data parallel training approach, and, hence, work better with horizontal scaling. Time-series forecasting problems are less amenable, and, hence, require more vertical scaling.

We have not addressed model parallel training. Although other distributed deep learning frameworks~\cite{horovod18,dai2019bigdl} do the same, model parallelism is desirable when big models with 1 billion parameters or more do not fit in memory. To implement model parallelism, we will need to handle more dependencies among subtasks than we did with data parallelism. 

\input{discussion}

\bibliographystyle{IEEEtran}
\bibliography{IEEEabrv,eydle-ieee}

\listoffixmes

\end{document}

%% file: intro.tex
\section{Introduction} \label{sec:intro}

The field of Deep Learning (DL) is growing rapidly. Its commercial applications include computer vision, natural language processing, sentiment analysis and speech synthesis. DL is widely used in processing social media data being generated at a staggering pace -- over 600 million tweets generated daily, billions of images uploaded to Facebook every day and over 400 hours of video content uploaded to YouTube per minute. 

The size of DL models used in the industry has grown from millions to billions to trillions of parameters. The size of training datasets has grown to billions and trillions of samples~\cite{horovod18,petuum,turingnlg,googleswitch2021}. We need immense computing power to train big models over big data, and its cost can be prohibitive. The cost of training a single model can range from hundreds to thousands of dollars or more~\cite{fakenews}. Training many models or continually retraining a model can add to the cost. Furthermore, training models to acceptable levels of accuracy can take weeks~\cite{fakenews}. To reduce training time, the industry needs distributed training solutions~\cite{horovod18}. Improvements in computing hardware such as GPU and TPU, and techniques such as transfer learning~\cite{Pan2010TransferLearning} have reduced training time and cost for some DL applications. Other applications -- creating new deep learning models in novel areas where the model architecture is unknown; hyperparameter tuning and neural architecture search for finding optimal device-specific models; and training an existing model on large amount of new task-specific data -- need solutions to reduce training time and cost.  

The cost of training DL models using cloud computing or dedicated CPU/GPU cluster computing can be prohibitive for small- and mid-sized businesses, and academic researchers. We explore training DL models using an alternative computing paradigm called Volunteer Computing (VC), which can provide petaFLOPS to exaFLOPS distributed computing power at low cost~\cite{kondo2009cost}. Applications designed to run on VC systems have to handle three main characteristics: \textit{fault tolerance}, \textit{variable network latency} and \textit{heterogeneity} of computing devices. Current DL training solutions are primarily built for cluster computing systems and require low latency, high throughput, high reliability and homogeneous nodes~\cite{distbelief,adam,horovod18,dai2019bigdl}. Distributed versions of popular deep 
learning frameworks such as tf.distribute, torch.distributed and Horovod run predominantly on cluster computing systems. In contrast, we design a distributed DL training system that can handle the three main characteristics of VC systems. Our aim is to provide a distributed solution that can achieve acceptable levels of accuracy and training time, and lower computing costs.   

Traditional VC systems provide massive computing at low cost by offering non-monetary incentives to volunteers who donate their idle device resources to the VC system. Since computing happens on arbitrary volunteer devices, data security can be an issue for commercial applications that have stringent requirements on data sharing. To address this scenario, we leverage a feature of 
commercial cloud platforms called \textit{preemptible instances}.\footnote{\url{		https://cloud.google.com/compute/docs/instances/preemptible}} Commercial cloud platforms such as Google Cloud and Amazon Web Services (AWS) provide more robust data security. Preemptible instances cost 70-90\% less than the standard computing instances, but they can be terminated by the cloud provider at any time. Only applications that are fault tolerant can work with preemptible instances. Because our system is fault tolerant, we can use preemptible instances to lower cost and improve data security.  

In Section~\ref{sec:background}, we elaborate on the limited work~\cite{jsdoop,kijsipongse2018hybrid,desell2017developing,ryabinin2020crowdsourced} that exists on running distributed DL training using a VC-like paradigm. 
Our main contributions are as follows:
\begin{itemize}
 \item We design a distributed DL system that can run on a VC-like paradigm, and handle fault tolerance, variable network latency and heterogeneity of computing nodes.  
 \item We lower cost and improve data security of distributed DL training by leveraging preemptible computing instances in commercial cloud platforms.
\item We implement VC-ASGD, a novel asynchronous stochastic gradient descent (ASGD) optimization scheme designed for distributed DL training using the VC paradigm.  
\item We show how to improve scalability of distributed DL training by using an eventual consistency database.
\end{itemize} 

We organize the rest of the paper as follows: background and related work in Section~\ref{sec:background}; system design in Section~\ref{sec:sysdesign}; experiments and results in Section~\ref{sec:results}; limitations in Section~\ref{sec:limitations}; and summary and conclusions in Section~\ref{sec:discussion}.

%% file: background.tex
\section{Background and Related Work} \label{sec:background}

In this section, we provide background required to understand this work
and discuss closely related work.

\subsection{Volunteer Computing} \label{sec:vcsystems}
\input{vc}

\subsection{Distributed Deep Learning} \label{sec:distdl}
As DL datasets and models have grown in size, researchers have explored scaling up 
DL training through distributed computing. A distributed DL strategy needs to efficiently divide a DL training job into subtasks, distribute them over multiple nodes, coordinate any dependencies, and aggregate the results. Current strategies for distributed DL are primarily designed for cloud and dedicated computing clusters. These include strategies from popular DL frameworks such as TensorFlow (tf.distribute) from Google, PyTorch (torch.distributed) from Facebook, BigDL from Intel and Horovod from Uber. The strategies use paradigms such as Bulk Synchronous Parallelism (BSP) or variants such as MapReduce, AllReduce and RingReduce, Stale Synchronous Parallelism (SSP), Message Passing Interface (MPI), Graph Dataflow, and Parameter Server approach. They require one or more of the following: (a) equal high-throughput communication among the compute nodes; (b) synchronous communication among the nodes; (c) peer-to-peer communication among nodes; (d) fault tolerant nodes; (e) fixed node IP addresses and ports; and (f) administrative access to the nodes. However, such requirements cannot be supported in a system operating using a VC-like paradigm. Several cloud services providers such as IBM, Google and Microsoft provide deep learning as a service (DLaaS) where users can upload DL models and code to the cloud, and train the models without having to manage the DL training infrastructure~\cite{bhattacharjee2017ibm}. The primary goal for DLaaS is usability and not necessarily cost savings or speed of training. Internally, providers may implement features such as fault tolerance, but the features are designed for the cloud paradigm and not the VC paradigm.  

Distributed training can use \textit{synchronous} or \textit{asynchronous} strategy. In a synchronous strategy, compute nodes may need to share a clock, and/or they may have to be tightly connected to each other via high speed communication channels such as InfiniBand or Gigabit Ethernet switch. Synchronous strategies such as MapReduce, AllReduce or RingReduce for model training require all compute nodes to operate in sync. An asynchronous strategy does not require nodes to be in sync and is better suited for VC systems. Stochastic Gradient Descent (SGD) is a widely used scheme to update model parameters, also known as weights, during iterative training of the models~\cite{bottou-curtis-nocedal2018_SGD}. There are synchronous and asynchronous variants of SGD. In an Asynchronous SGD (ASGD) scheme, clients train on a copy of the model using their assigned dataset and send their parameter updates, either gradients or weights, asynchronously to the parameter server. The parameter server stores the central copy of the model parameters, updates it, and sends it to the clients when needed~\cite{LiParamServer_2014}. Here, \textit{gradient} refers to the gradient of the loss function of the neural network with respect to the model parameters. 

Google's DistBelief framework proposed~\textit{Downpour} SGD as a distributed training scheme that requires clients to use high frequency, high bandwidth parameter synchronization with the parameter server~\cite{distbelief}. Clients send their gradients every $n_{push}$ training iterations and receive the updated server parameter copy every $n_{fetch}$ iterations, where $n_{push}$ and $n_{fetch}$ are communication frequency parameters. This type of communication is suited for a homogeneous cluster with clients that can maintain their state throughout the training job, but it is not suited for the VC environment. Downpour SGD distributes a large DL job using a data parallel approach; each client has a local copy of the entire model, and this local model is trained using a smaller subset of the training data. Thus, if there are $n$ clients, there are $n$ independent copies of the model and $n$ local copies of the model parameters. There are one or more parameter servers that receive clients' local parameters and update the centrally stored parameters using lock-free data structures similar to Hogwild!~\cite{hogwild}. Microsoft Adam shares similarities with Downpour SGD but has additional enhancements for reducing communication and computation overheads~\cite{adam}. Asynchronous training using the parameter server strategy in Google's Tensorflow is not fault tolerant against failure of  the centralized server~\cite{TensorflowOSDI2016}. This strategy, which is currently limited to CPUs, does not allow changing the number of clients and parameter servers dynamically. 

Petuum is a distributed machine learning platform for training big models on big data using data- and model-parallel approaches~\cite{petuum}. It uses a Stale Synchronous Parallel approach to assimilate parameter updates from clients. It exploits the error tolerance and uneven convergence properties of the SGD optimization algorithm and the dependency between model parameters to train faster. 
Petuum schedules work on the clients and server such that low dependency and non-converged parameters get higher priority than other parameters. 

Asynchronous training schemes are prone to the problem of~\textit{delayed gradients} or staleness of the parameters~\cite{distbelief,HoStaleSynchronous2013,Lian2015DelayedGradient,zhang2015}, which leads to slower training and lower accuracy at the end of training compared to a synchronous training scheme. The frequency of client-server communication, e.g. values of $n_{push}$ and $n_{fetch}$ in Downpour, can be tuned to reduce the network communication overhead while maintaining the freshness of the server and client parameter copies, although less frequent communication has been reported to slow down training.  
In the Elastic Averaging ASGD (EASGD) method~\cite{zhang2015}, the authors proposed to reduce the communication overhead between clients and server by using local clocks on them and using a single communication frequency parameter to control the frequency of sending and receiving parameter updates between the clients and server. The communication is still based on a homogeneous cluster paradigm: a client sends a request for the server parameter copy, waits for the server to send it,  computes the difference between client's and server's parameter copies, and sends back the difference to the server, which applies the difference to its parameter copy. 

To alleviate the delayed gradient problem, while maintaining the asynchronicity of the distributed SGD algorithms, different approaches have been used. Downpour used \textit{warmstarting}, where serial synchronous training is performed for multiple epochs before starting distributed training. Petuum bounds the staleness of the parameters to address the delayed gradient problem. Zheng et al.~\cite{zheng2017} proposed~\textit{Delay Compensated ASGD} or DC-ASGD. Here, a computationally cheaper approximation of the Hessian matrix (of the loss function with respect to the parameters) is constructed at the server in terms of client's gradient and client's pre-training parameter copy. The Hessian approximator is meant to increase the rate of convergence of the ASGD method from first-order to second-order and thereby compensate for the accuracy loss due to delay. However, similar to the ASGD schemes discussed above, the DC-ASGD scheme needs parameter updates from all clients that are training on individual subsets of the data and, hence, is not fault tolerant. 

\subsection{BOINC} \label{sec:boinc}

The Berkeley Open Infrastructure for Network Computing (BOINC) project from the University of California Berkeley develops a free open source middleware software for building VC systems~\cite{boinc}. It also runs a VC system with 791,000 volunteer computers. The system runs distributed applications in areas such as mathematics, linguistics, medicine and environmental science. BOINC software is used by several universities, research labs and organizations. Notably, the World Community Grid\footnote{https://www.worldcommunitygrid.org/} VC system from IBM uses BOINC. We build our system on top of the BOINC middleware software.

BOINC uses client server architecture. At a high level, BOINC consists of server and client components. A BOINC server can host many projects each with multiple applications. An application can run jobs with multiple subtasks, which are called workunits in BOINC terminology. A BOINC client software runs on volunteer devices. The BOINC server has a \textit{scheduler} service that sends workunits to the clients and tracks their progress. BOINC server also supports a \textit{validator} service, which determines if the results of a job are valid, and an \textit{assimilator} service, which  processes the results. Communication between BOINC server and client happens over HTTP(S) protocol. BOINC uses a database server to track information about clients, workunits and results. It uses a web server to distribute application code and data. 

BOINC client code can run on many operating systems including Microsoft Windows, MacOS, Linux and Android. The application code can be built for each operating system or can be run agnostic to the operating system using virtualization. Applications can be built using the BOINC API. Legacy programs or programs written in languages that do not require compilation, such as Python, can be run using BOINC \textit{wrappers} for specific operating systems. BOINC scheduler can assign workunits to clients based on the client devices' resources such as CPU, GPU and RAM. 

Volunteer computers may join or leave projects at will, and users may 
start or shutdown their devices any time. To handle this, BOINC tracks the 
status of workunits. If the result of a workunit is not received within a configurable time limit, it is  rescheduled to run on another client. BOINC allows a workunit to be replicated on multiple clients to create computational redundancy, which can help with fault tolerance and verification of results.

%% file: vc.tex
In the Volunteer Computing (VC) paradigm, volunteers donate idle computing power 
of their devices to VC systems. VC systems can be grid-based or browser-based. Grid-based systems require volunteers to download and install client software on their devices. Older browser-based systems required volunteers to download and install a browser plug-in, but, in newer systems, users visit a web address from their browser. In this work, we focus on grid-based systems.

VC can provide massively distributed computing power at low cost because volunteers donate their idle computing resources for altruistic and not monetary incentives. Because of increased interest in VC due to the COVID-19 pandemic, the Folding@home\footnote{\url{https://foldingathome.org/}} VC system reached 2.43 exa Floating Point Operations Per Second (FLOPS) in April 2020 becoming the world's first exaFLOPS system~\cite{foldingathome}. The Berkeley Open 
Infrastructure for Network Computing (BOINC) VC system can achieve a 24-hour average of 
41.58 petaFLOPS using 791,443 volunteer computers~\cite{boinc}.  Kondo et al. estimated that 0.1 peta FLOPS of computing costs \$125,000 on BOINC compared to \$175 million on AWS cloud platform~\cite{kondo2009cost}. 

Existing distributed deep learning systems are primarily built for cluster  
computing systems. VC systems have different characteristics compared to cluster computing 
systems. The latter have low latency, high throughput, high reliability and 
homogeneous nodes. They generally support synchronous and/or peer-to-peer 
communication among nodes and provide administrative access to the nodes. VC 
systems can have higher latency and lower throughput because volunteer nodes 
can connect via lower speed Internet connections (WAN) instead of high-speed 
local network connections (LAN). They have lower reliability because nodes can 
connect and disconnect any time. VC nodes include heterogeneous devices such as 
desktops, laptops and tablets, and asynchronous communication among the nodes 
is more realistic than synchronous communication. Peer-to-peer communication 
among nodes is difficult because of the client server environment. Lastly, 
administrative access to client nodes is not guaranteed. Traditional VC systems run on arbitrary volunteer nodes that cannot be trusted. This can be a drawback for commercial applications that require strong guarantees on data security.  

Little work exists on building a solution to run distributed DL training on VC systems~\cite{jsdoop, kijsipongse2018hybrid, ryabinin2020crowdsourced}. Morell et al. built JSDoop~\cite{jsdoop}, a browser-based VC system for training an RNN model to predict text. Our work uses grid-based VC systems. Kijsipongse et al. explore combining cluster and VC systems for DL~\cite{kijsipongse2018hybrid}. In contrast, we explore DL in a VC-like system, and design a system that can handle fault tolerance, variable network latency and heterogeneity of nodes. Our design enables us to run distributed DL training using preemptible instances on commercial cloud platforms. This lowers the costs by 70-90\% compared to training with standard computing instances on cloud platforms. Furthermore, it provides stronger security guarantees than traditional VC systems. Ryabinin and Gusev propose to train large neural network models  using a decentralized mixture-of-experts (MoE) method that splits the  network into smaller sub-networks, which train on volunteer devices~\cite{ryabinin2020crowdsourced}. The MoE approach is closer to a model parallel approach than to our data parallel approach.

The Machine Learning Dataset Generator (MLDS) project\footnote{https://www.mlcathome.org/} is building a dataset consisting of thousands of neural networks trained on similar, highly controlled data. The MLDS project runs on the BOINC VC system. Each neural network in the dataset is obtained by running a training job on a volunteer computing device. Desell used volunteer computers from a BOINC VC system to train CNNs for a neuro-evolution algorithm~\cite{desell2017developing}. Unlike our work, the MLDS project and work by Desell do not use data parallel training approach to split a training job into multiple training subtasks.  


%% file: sysdesign.tex
\section{System Design} \label{sec:sysdesign}
\begin{figure}
\centering
 \includegraphics[scale=0.5]{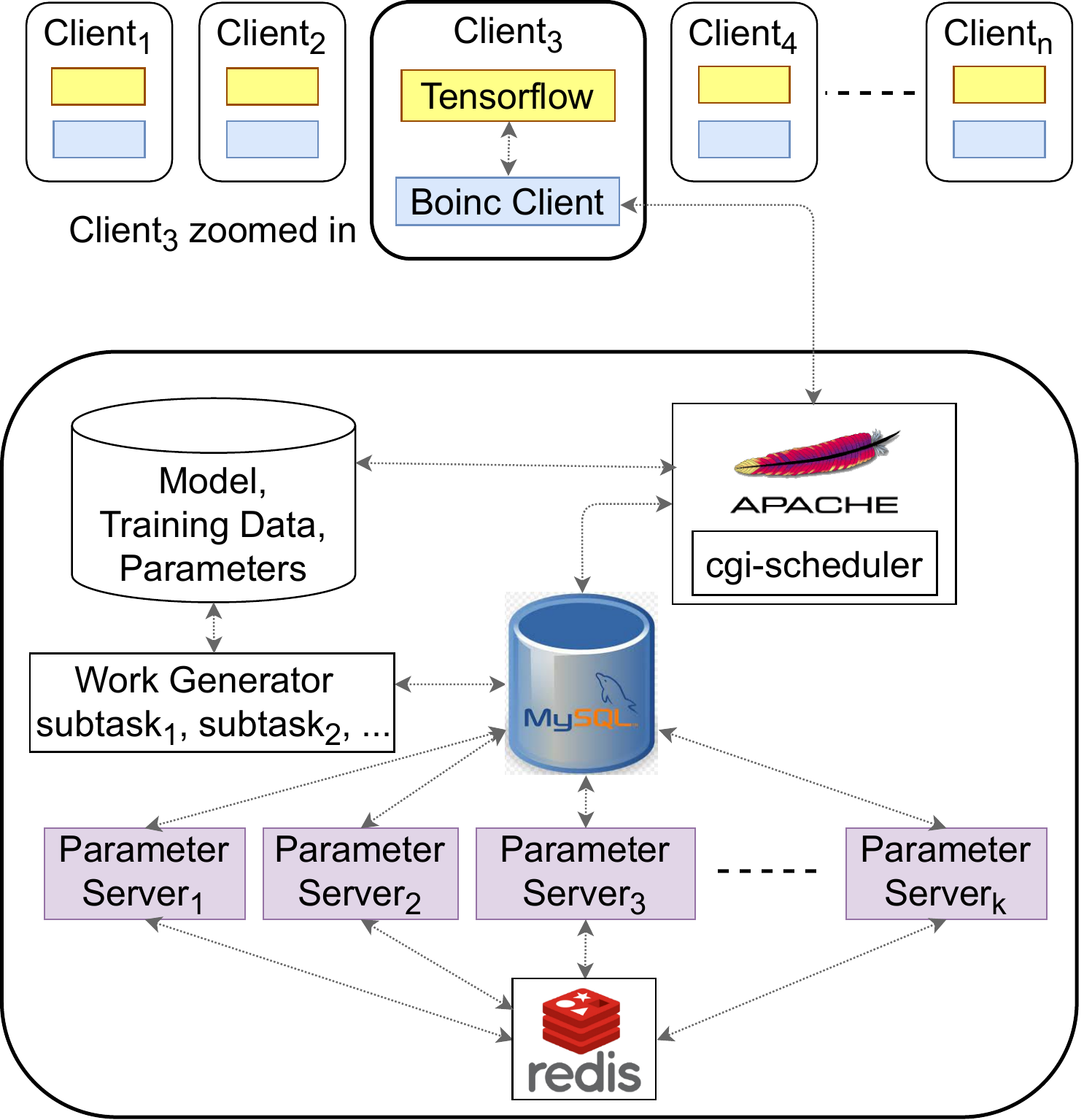}
 \caption{Main Components of our System} 
\label{fig:sysdesign}
\end{figure}

In this section, we discuss key design decisions for building a distributed deep learning system that can run on VC-like systems. We highlight design decisions for handling fault tolerance, latency, heterogeneity, scalability and security. Figure~\ref{fig:sysdesign} shows the main components of our system.

We use the Berkeley Open Infrastructure for Network Computing (BOINC) middleware software~\cite{boinc} for building our distributed deep learning system. We use BOINC because it is a popular, free, open source software with active community of developers and users. Section~\ref{sec:boinc} describes BOINC software, its components and how they support building a VC system. 

Traditional VC systems implemented using BOINC perform well with applications that execute as Embarrassingly Parallel tasks where there is little dependency and communication among the tasks. However, when a distributed DL training strategy splits a DL job into multiple training subtasks, there are dependencies among the subtasks. In subsequent sections, we detail how we adapt BOINC to run DL training jobs. 

\subsection{Implementing Distributed Training} \label{sec:design1}
BOINC uses client-server architecture. 
Clients contact the server for workunits, execute the workunits and upload the results back to the server. To implement distributed DL model training over a client server architecture, we use \textit{data parallel} training with \textit{parameter server} approach explained in Section~\ref{sec:distdl}. Our parameter server is built on top of BOINC's configurable~\textit{assimilator process}. The \textit{work generator} component splits a single DL training job into multiple training subtasks. In BOINC, each training subtask maps to a workunit. To create subtasks, the work generator splits the DL training dataset into subsets. For instance, if the work generator splits the training dataset into 50 subsets, it creates 50 training subtasks. In addition to a data subset, a subtask also contains a model with layer architecture, a copy of model parameters, and training code to be run on the client. One epoch consists of training over the entire dataset, and an epoch is over when subtasks training over the data subsets of the dataset are complete. 

A user running a training job has to specify details such as the model, dataset and accuracy. However, the design of the work generator automatically handles the details of converting a training job into a data parallel training job. This entails deciding the best possible split for the training dataset, creating the training subtasks and running multiple epochs of training until a stopping criterion is met. In the past, parameter server and data parallel approaches have been difficult to use because users have to deal with the details~\cite{horovod18} of running data parallel training. Our design tries to overcome this hurdle.

A client receives one or more subtasks when it sends a request to the scheduler. The client downloads model, parameter and data files for a subtask from the BOINC web server, and then trains the model on the data using Tensorflow library. After training is complete, it uploads the model parameters to the BOINC web server. As part of processing the results of the subtask, BOINC invokes the parameter server, which uses a distributed parameter update scheme to combine model parameters received from the subtask. We implement a novel parameter update scheme called VC-ASGD. After assimilating a parameter update from a training subtask, the parameter server computes the validation accuracy. At the end of an epoch, the parameter server calculates the average validation accuracy over all the subtasks. If the average validation accuracy meets the required accuracy threshold, the training stops. If not, the training proceeds with the next epoch.  

\subsection{Handling Fault Tolerance, Latency and Heterogeneity} \label{sec:faulttol}
If a client executing a training subtask terminates unexpectedly or the network communication fails, the result of a training subtask will not reach the parameter server. To make the training fault tolerant, we use BOINC's scheduler feature. If the results from a client do not arrive within a configurable time limit, the scheduler reassigns the training subtask to another client. The scheduler can track how reliably clients return results and assign subtasks to more reliable clients.  

In our system, model parameter updates from clients can arrive at the parameter server at different times due to three reasons. First, heterogeneous clients can execute training subtasks at different speeds. Second, since clients can be in different geographical regions, they can communicate with the parameter sever with variable network latency. Lastly, if a training subtask is rescheduled because of a faulty client, the parameter update for that subtask can be delayed. To address the fact that parameter updates can arrive at different times, we implement an asynchronous distributed parameter update scheme called VC-ASGD. It is asynchronous because the parameter server does not wait for parameter updates from all subtasks before updating the server parameter copy. 

We use compression and caching to minimize network latency in transferring model, data, parameter and code files. For DL problems such as image classification, training data can be large. When the files are not in compressed formats such as \textit{.npz} or \textit{.h5}, we can use a BOINC feature that automatically compresses a file on the server and decompresses it on the client. We also use the BOINC \textit{sticky-file} feature to cache model, data and code files on a client instance. If a client has a cache of a training data file, in order to avoid multiple downloads, the BOINC scheduler tries to assign subsequent training subtasks involving that file to that client.  

\subsection{Implementing Asynchronous Parameter Updates} \label{sec:vcasgd}
In Section~\ref{sec:distdl}, we discussed asynchronous training as a popular method for distributed  training. We discussed schemes such as Downpour SGD, asynchronous Elastic Averaging SGD (EASGD) and Delay Compensated ASGD. Using Downpour SGD as-is can lead to consistent loss of updates from a slow or disconnected client leading to suboptimal training. EASGD requires updates from all clients, which can cause significant delay at the server in waiting for the slow or disconnected clients, and hence, is not fault tolerant. Some of the methods also require maintaining local clocks on clients, which can not be ensured in a VC-like environment. Therefore, a new asynchronous parameter update scheme is sought. Here, we present VC-ASGD, an asynchronous parameter update scheme that is convergent in a VC-like environment and does not impose the above requirements.

As discussed earlier, the parameter server receives parameter updates from clients that are executing training subtasks. When the parameter server receives an update, it immediately assimilates the received parameters with the server copy, regardless of the order in which updates are received. Because the parameter server does not wait for updates from all subtasks, the scheme is fault tolerant. The update at the parameter server can be represented by the following equation:
\begin{equation} \label{eq:ddl}
	W_s \leftarrow \alpha W_s + (1 - \alpha) W_{c_i,j}
\end{equation}
Here, $W_s$ denotes the server parameter copy, $W_{c_i,j}$  denotes the parameter copy received from a client $i$ after executing training subtask $j$, and $\alpha$ is the VC-ASGD hyperparameter. The server update calculations remain opaque to the clients, and clients work independently on the assigned training subtasks using a copy of the server parameter $W_s$ sent along with the subtasks. If there are $n_t$ training subtasks in an epoch, and all of them return results to the server, the server applies Equation~(\ref{eq:ddl}) $n_t$ times. Using Equation~(\ref{eq:ddl}) recursively over $n_t$ returning subtasks, the server parameter $W_{s,e}$ at the end of epoch $e$  can be
 expressed in terms of the server parameter $W_{s,e-1}$ at the end of epoch $e-1$ as follows, 
\begin{equation} \label{eq:ddl1}
	W_{s,e} = \alpha^{n_t} W_{s,e-1} + (1 - \alpha) \sum_{j=1}^{n_t} \alpha^{n_t-j} W_{c,j}
\end{equation}
This shows how $\alpha$ controls convergence of the model ($W_{s,e-1}$ approaching $W_{s,e}$ as $e$ increases) by modulating the impact of training at clients, which is captured in the $W_c$ term. The summation over subtasks happens asynchronously and is fault tolerant. 
For faster convergence, we allow $\alpha$ to vary with the epoch number $e$. Motivated from the convergence analysis of the SGD algorithm~\cite{pmlr-v48-hardt16}, we explore a special case of $\alpha$ increasing with $e$ in Section~\ref{sec:results}. This is analogous to the \textit{learning rate scheduler} used in optimizers such as SGD~\cite{distbelief}.

\subsection{Improving Scalability} \label{sec:concurrency}
To increase the speed of distributed DL training, we can use more clients. As we scale up the number of clients, a single parameter server can become a performance bottleneck. However, if we use multiple parameter servers, we need a mechanism through which these servers can concurrently access a shared copy of server parameters. To address this, we store a copy of the server parameters in a database. Other approaches include file-locking and shared memory updates. Parameter servers accessing a shared file stored on a network file system can slow down the parameter updates. A shared memory approach can be faster, but the parameter servers have to be implemented as processes on a single server, which prevents us from scaling horizontally.  

The choice of database can impact the speed of parameter updates. A traditional relational database using \textit{strong consistency} will apply concurrent parameter updates to the shared copy of the server parameters in a serializable order. However, strong consistency can lower scalability. An \textit{eventual consistency} database improves scalability, but can lose some parameter updates. Prior work shows that distributed training can tolerate loss of some parameter updates without a significant impact on the accuracy of training~\cite{distbelief,petuum,adam}. Hence, we use an eventual consistency database in our design.

Main-memory databases store as much data in memory as possible to avoid disk I/O latency. Availability of robust computing hardware and high capacity RAMs have made main-memory databases popular. To handle concurrent parameter updates, we use Redis,\footnote{\url{https://redis.io/}} which is a main-memory eventual consistency database. Redis stores data as key value pairs. We store all the parameters of a model as a single value.  

Recall that a client uploads the parameter update from a training subtask to the BOINC web server, and BOINC invokes the parameter server for processing the update. In our current design, BOINC evenly distributes the load to multiple parameter servers. Only one parameter server processes the update from a training subtask. Prior work has shown that users find it difficult to determine the ratio of the number of parameter servers to the number of clients~\cite{horovod18}. Hence, our idea is to allow the system to dynamically vary the number of parameter servers based on the number of jobs and clients. 

\subsection{Enhancing Data Security} \label{sec:vcsec}
Traditional VC systems run workunits on volunteer client devices. The computation cost of VC systems is low because volunteers are generally not provided monetary compensation. However, volunteer devices cannot be trusted to provide strong guarantees on data security, which can be an issue for commercial applications. To address this, we leverage a feature of commercial cloud platforms called \textit{preemptible instances}, which cost 70-90\% less than the standard computing instances because they come from unused excess capacity. The cloud provider can terminate them at any time.
	
Commercial cloud platforms such as the Google Cloud and AWS provide data security guarantees, but their standard computing instances drive up the cost of DL training. Preemptible instances can drive down the cost, but applications have to be fault tolerant to use them. Our system is designed to handle fault-tolerance and can use preemptible instances to lower cost and improve data security. In our design, each training subtask is assigned a task completion timeout period. If a client running on a preemptible instance fails to return the result of a subtask because of instance termination, the subtask is reassigned to another instance after the timeout period. 

We run clients on a fleet of preemptible instances. Because our system can handle heterogeneous devices and network latency, we can lower the cost further by using different types of instances as well as instances running in different data centers and geographical regions.

%% file: expt.tex
\section{Experiment Design and Results} \label{sec:results}
Here we present the experimental design for validating our system design and discuss results from our experiments. 

\subsection{Experimental Setup}
We run our experiments on the AWS cloud platform. For the server infrastructure, we use a single standard computing instance. On this instance, we run all the parameter servers, Redis database server, BOINC Apache web server and BOINC MySQL database server. We use a fleet of computing instances of different types for running the clients. Each instance runs one client. Depending on the experiment, we use either standard or preemptible instances. Table~\ref{tab:instance-config} shows the configuration of server and client instances. The server instance runs Ubuntu OS, and the client instances run Ubuntu or Mac OS.

\begin{table}[t]
	\centering
	\begin{tabular}{@{}ccccc@{}}
		\toprule
		& vCPU & \begin{tabular}[c]{@{}c@{}}Clock\\ Speed \\ (GHz)\end{tabular} & \begin{tabular}[c]{@{}c@{}}RAM \\ (GB)\end{tabular} & \begin{tabular}[c]{@{}c@{}}Network\\ Bandwidth\\ (Gbps)\end{tabular} \\ \midrule
		Server & 8 & 2.3 & 61 & upto 10 \\
		Client & 8 & 2.2 & 32 & upto 5 \\
		Client & 8 & 2.5 & 32 & upto 5 \\
		Client & 8 & 2.8 & 15 & upto 2 \\
		Client & 16 & 2.8 & 30 & upto 2 \\ \bottomrule
	\end{tabular}
	\caption{Server and client instance configurations}
	\label{tab:instance-config}
\end{table}

We use the CIFAR10 image classification problem~\cite{Sermanet2014Cifar10} for benchmarking training accuracy and time. The CIFAR10 dataset consists of 60,000 32$\times$32 color images in 10 classes with 6,000 images per class, and is split into 50,000 training and 10,000 test images. Using a data parallel approach, we split the training dataset into 50 subsets. Each subset is stored in compressed \textit{.npz} format and is 3.9MB in size. In one training epoch, we train over the 50 data subsets using 50 training subtasks. We use the ResNetV2 model~\cite{He2016Resnet} with 552 layers, 4,972,746 total parameters and 4,941,578 trainable parameters. The model file is in \textit{.json} format and is 269KB in size. We store parameters in a compressed \textit{.h5} file; each parameter file is 21.2MB in size. We use a He-normal initializer to initialize the parameters randomly. We train the model implemented in Tensorflow using \textit{Adam} optimizer with a constant learning rate value of 0.001. We do not use momentum. To keep our model simple, we also do not use regularization and dropout techniques, which can improve generalization of a model by reducing overfitting to the training data. Our focus is on comparing different training strategies, and because we use the same model for comparison, these model-specific design choices do not affect our conclusions.

\begin{figure}[t]
	\centering
	\includegraphics[scale=0.35]{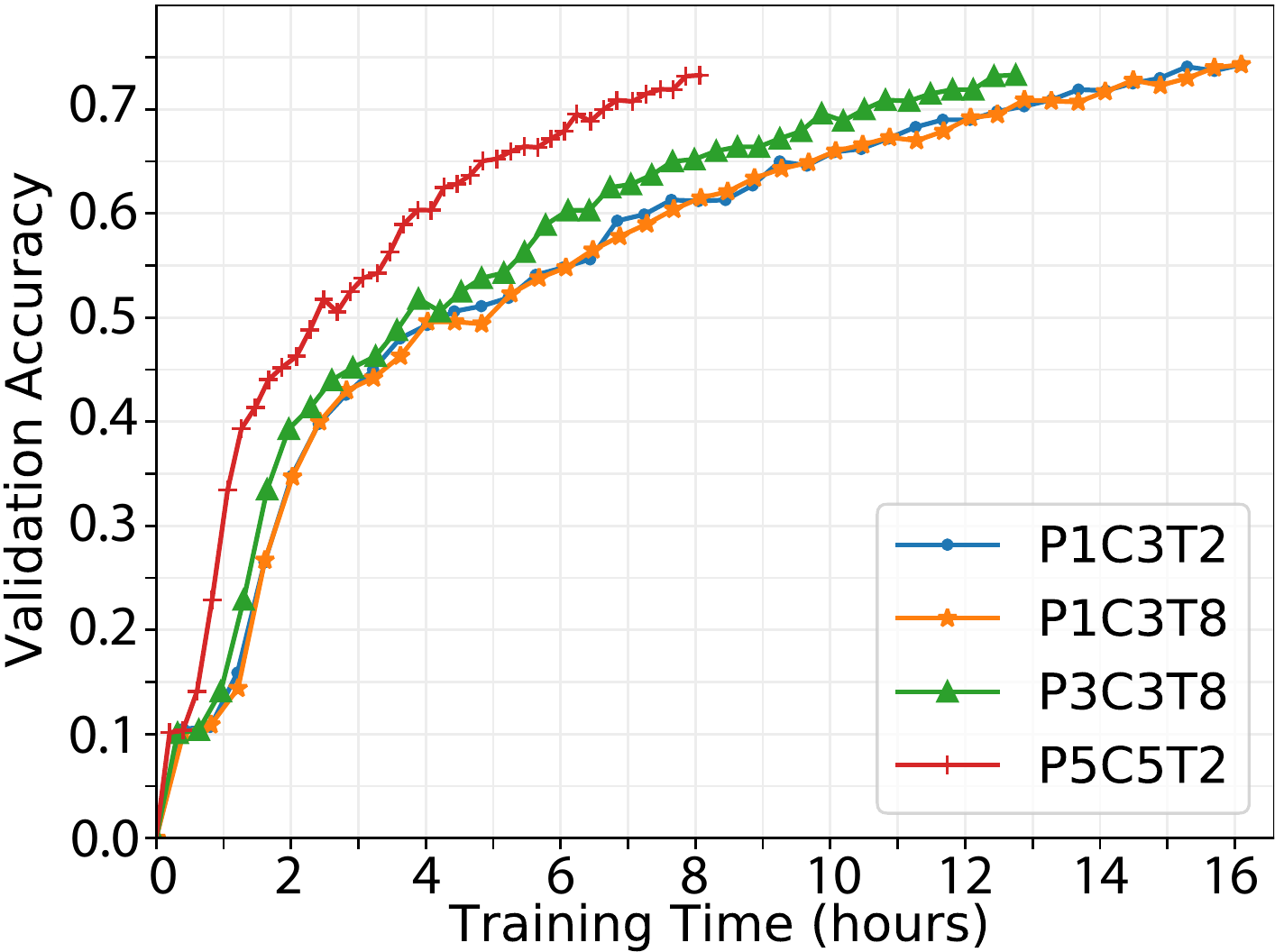}
	\caption{Effect of distributed training keeping $\alpha \!=\! 0.95$} 
	\label{fig:samealpha}
\end{figure}

We use the following abbreviations in discussing the results of our experiments. \textbf{P} denotes a parameter server, \textbf{C} denotes a client and  \textbf{T} denotes a training subtask assigned to a client. The number \textbf{n} in  P\textbf{n} and C\textbf{n} denotes the total number of parameter servers and clients respectively used in a training job. The number \textbf{n} in T\textbf{n} indicates the maximum number of subtasks that can run simultaneously on a client. For example, the curve P5C5T2 in Figure~\ref{fig:samealpha} indicates training using 5 parameter servers (P=5), 5 clients (C=5) and each client running maximum of 2 training subtasks at a time (T=2). 

\subsection{Impact of Distributed Training}\label{sec:disttrain}
To understand the impact of distributed training on training accuracy and time, we vary Pn, Cn and Tn while fixing the VC-ASGD hyperparameter to $\alpha \!=\! 0.95$. Figure~\ref{fig:samealpha} compares results from experiments P1C3T2, P1C3T8, P3C3T8 and P5C5T2. Markers on each curve represent results for an epoch $e$ of training. The y-axis represents the average validation accuracy for all training subtasks in $e$. The training time for $e$ is the total time taken to complete all the training subtasks in $e$, and corresponding validation and server parameter updates. The x-axis represents the cumulative training time until $e$. 

 Figure~\ref{fig:samealpha} shows that all distributed training experiments reach $\sim$0.73 accuracy, but some reach the value faster. This suggests that varying Pn, Cn and Tn impacts training time, but not the final training accuracy. The differences in training times are influenced by three main factors: (a) total time taken by the clients to process the training subtasks; (b) total time taken by the parameter servers to process the results received from the clients and (c) imbalance between client and server processing times. We can control these factors via \textit{vertical} and \textit{horizontal} scaling of client and server computing instances.

\begin{figure}[t]
	\centering
	\includegraphics[scale=0.35]{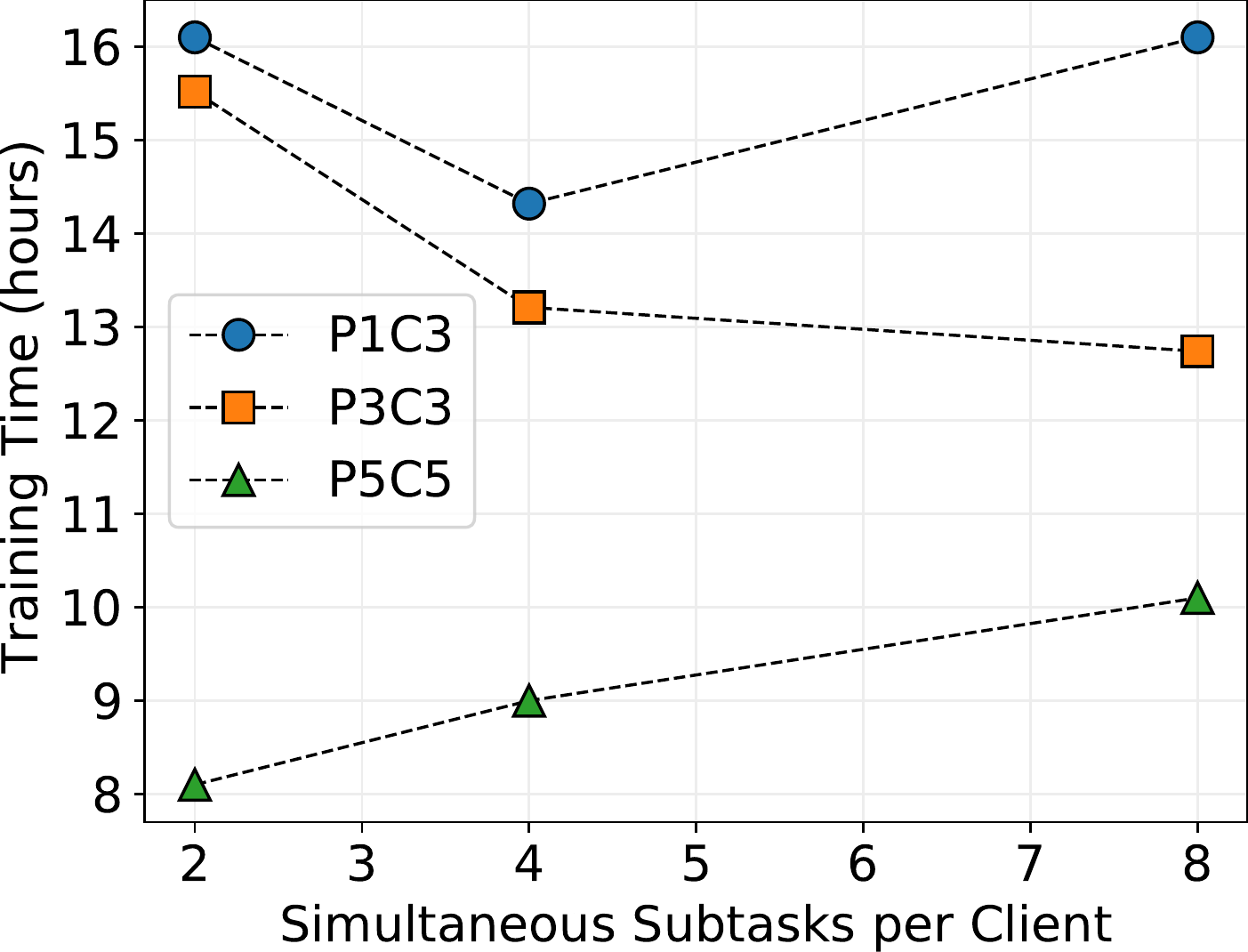}
	\caption{Effect of the number of parameter servers and the number of simultaneous subtasks on the training time. $\alpha 
		\!=\! 0.95$} \label{fig:jobsvtime}
\end{figure}

To illustrate the impact of the three factors listed above, in Figure~\ref{fig:jobsvtime}, we plot the training time for configurations P1C3, P3C3 and P5C5 for values T2, T4 and T8. Let us consider the imbalance between client and server processing times. The maximum number of subtasks that Pn has to assimilate at anytime is Cn $\!\times\!$ Tn. With P1C3, training time decreases from T2 to T4, but increases from T4 to T8. With T8, the three clients finish executing the subtasks much faster than a single parameter server can assimilate the results. To address this imbalance, we can increase the number of parameter servers running on the server computing instance. In P3C3T8, we increase Pn from 1 to 3, and the training time indeed decreases by 3 hours. With P5C5, the training time increases from T2 to T4 and T4 to T8. With increasing Tn, the imbalance between client and server processing times grows. We can reduce the imbalance by increasing Pn further.

When we increase Tn while keeping Cn constant, we scale vertically by running more subtasks on a single client. When we increase Cn, while keeping Tn constant, we scale horizontally by distributing subtasks to more clients. The throughput of the client computing instances in our experiments decreases after T8. The throughput of the server computing instance in our experimental setup decreases after P5. To reduce the total client and server processing times, we have to either use instances with more CPU and RAM, or use more instances. 

\subsection{Impact of VC-ASGD Hyperparameter}
\begin{figure}[t]
	\centering
	\includegraphics[scale=0.33]{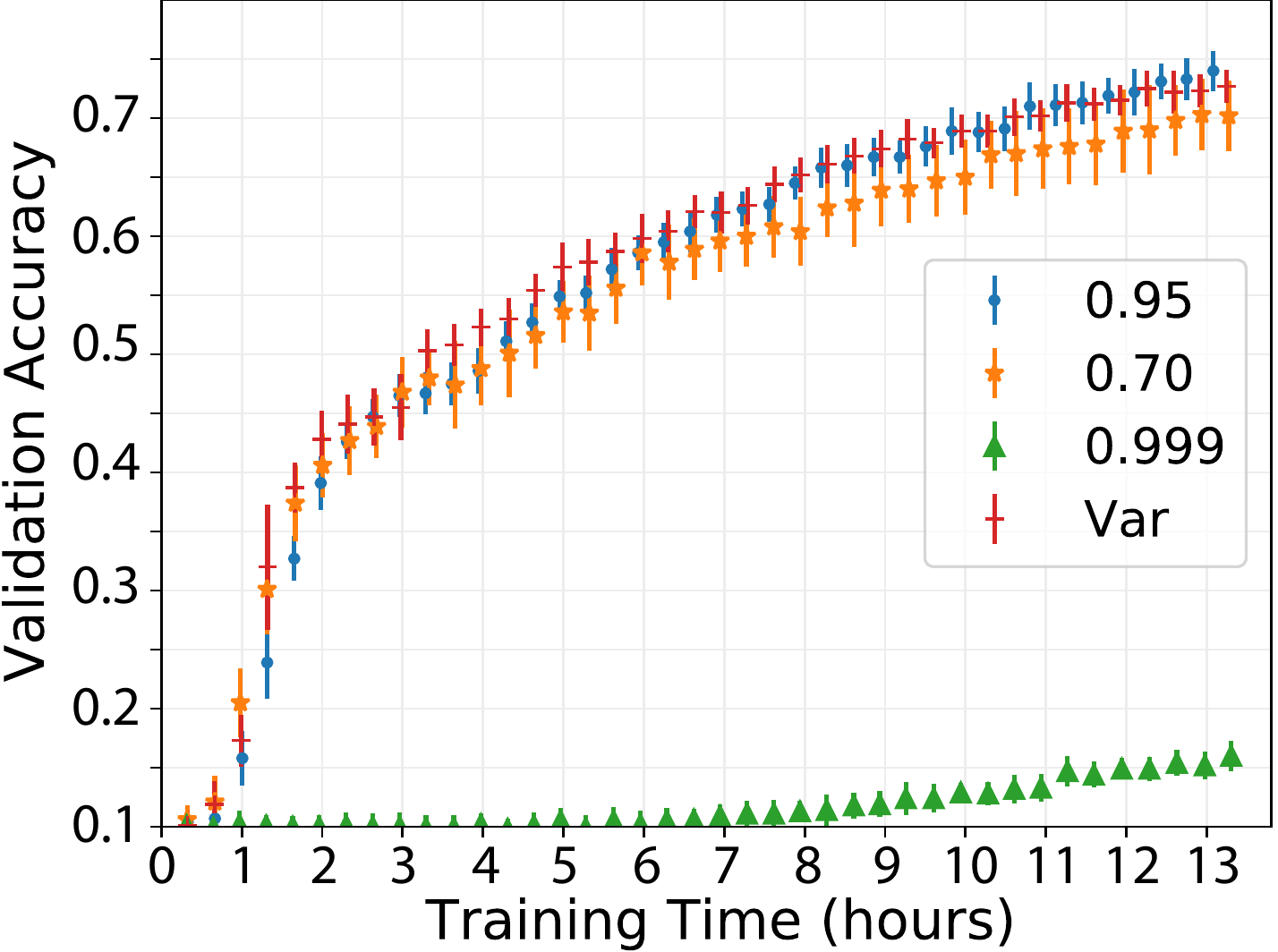}
	\caption{Effect of VC-ASGD hyperparameter $\alpha$ for 3 parameter servers, 3 clients and 4 simultaneous subtasks, i.e., P3C3T4} \label{fig:diffalph}
\end{figure}

In order to measure the effect of $\alpha$ on the validation accuracy across successive epochs, we conducted four different experiments with the P3C3T4 setup. We conducted three experiments with $\alpha$ values set to 0.7, 0.95 and 0.999 respectively. In the fourth experiment,  named \textit{Var}, we varied the $\alpha$ value as a function of the epoch number $e$. The results are plotted in Figure~\ref{fig:diffalph}. The error-bars in Figure~\ref{fig:diffalph} show the range of the accuracy values across 50 subtasks within each epoch, which can act as a proxy for the standard deviation of accuracy. Below we analyze the evolution of both the average accuracy and the standard deviation of accuracy per epoch.

For the first few epochs ($e \!<\! 7$), the average validation accuracy increases faster for the smaller $\alpha$ value of 0.7 than for 0.95. The reason is that the rate at which the server parameter copy \textit{learns} from clients is proportional to $(1 \!-\! \alpha)$ (see Equation~(\ref{eq:ddl1})), which is larger for smaller $\alpha$. In later epochs, the accuracy vs. time trend reverses such that $\alpha\!=\!0.95$ shows higher accuracy than $\alpha\!=\!0.7$. The reason is that, at any given epoch, clients are exposed to only subsets of the data, not the entire dataset. This slows down convergence of the client parameter copy to an optimum over the entire dataset. As the client is exposed to different subsets in different epochs, it is forced to \textit{unlearn} some of the features that it learned in previous epochs. This slows down learning of the features that are common over the entire dataset and, therefore, degrades generalization. As a result, a higher emphasis on partial learning of clients with $\alpha\!=\!0.7$, compared to $\alpha\!=\!0.95$, lowers the accuracy at the server in latter epochs. 

Next, we discuss evolution of the standard deviation of accuracy with time. The standard deviation of accuracy depends on the standard deviation of the server parameter copy, which increases proportional to the standard deviation of client parameter copy times the $(1 \!-\! \alpha)$ proportionality factor as per Equation~(\ref{eq:ddl1}). Hence, a smaller $\alpha$ means a larger standard deviation of accuracy. The standard deviation of client parameter copy is always positive because different clients are exposed to different data subsets.

We performed an experiment with $\alpha \!=\! 0.999$, which may be considered analogous to an EASGD training experiment with the moving rate set to 0.001 in~\cite{zhang2015}. The EASGD experiments were done on a GPU-cluster interconnected with InfiniBand and showed high accuracy values for this value of $\alpha$. However, in a VC-like environment, $\alpha\!=\!0.999$ results in much lower accuracy and slower training compared to the other two $\alpha$ values discussed above. This  supports our hypothesis in Section~\ref{sec:vcasgd} that existing ASGD schemes, designed for homogeneous cluster environments, cannot perform well in a VC-like environment because of the requirements they impose. In our VC-ASGD method, $\alpha\!=\! 0.999$ means only $0.1\%$ of the client parameters are used to update the server parameter copy, which is too low to achieve fast training. Another observation is that $\alpha\!=\! 0.999$ leads to the smallest standard deviation of accuracy among the four experiments in Figure~\ref{fig:diffalph}. The reason is the same as given above for explaining the standard deviations of $\alpha \!=\! 0.95$ and $0.7$.

\begin{figure}[t]
	\centering
	\includegraphics[scale=0.24]{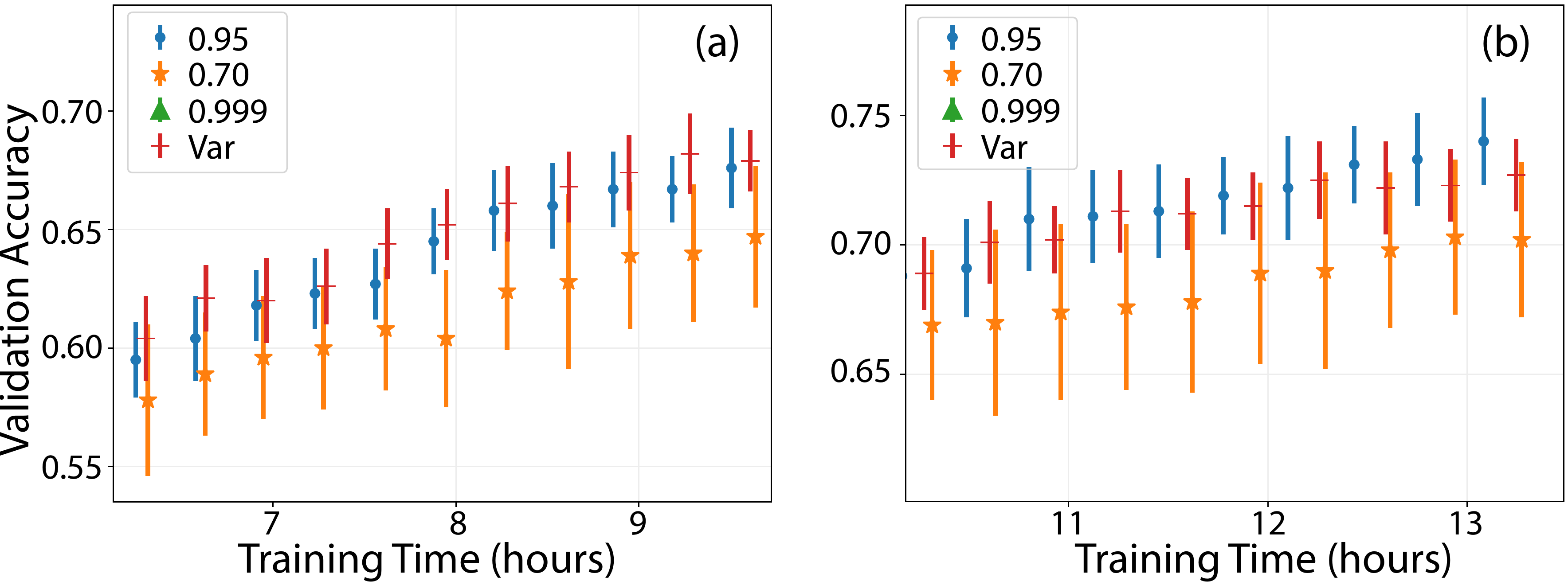}  
	\caption{Zoomed-in view of Fig.~\ref{fig:diffalph} between 6--10 hours (a) and near the end of training (b)} \label{fig:zoomin}
\end{figure}

The analysis of these three experiments conducted with different $\alpha$ values suggest that tuning of the hyperparameter $\alpha$ can yield faster training. It also suggests that changing $\alpha$ dynamically or adaptively with the epoch number can lead to faster training than keeping $\alpha$ constant. This is analogous to the concept of learning rate scheduling. To test this, we conduct an experiment with $\alpha_e \!=\! e/(e+1)$ such that $\alpha$ increases from $0.5$ to $0.98$ as the epoch number $e$  increases from 1 to 40. We observe that the accuracy increases at a faster pace than the $\alpha \!=\! 0.95$ case. Also, the standard deviation of accuracy is smaller than either the $\alpha \!=\! 0.7$ or $0.95$ case. This is illustrated in Figure~\ref{fig:zoomin} which shows zoom-in views of Figure~\ref{fig:diffalph} from 6--10 hour window and 10--14 hr window during training. 

\begin{figure}[t]
	\centering        
	\includegraphics[scale=0.29]{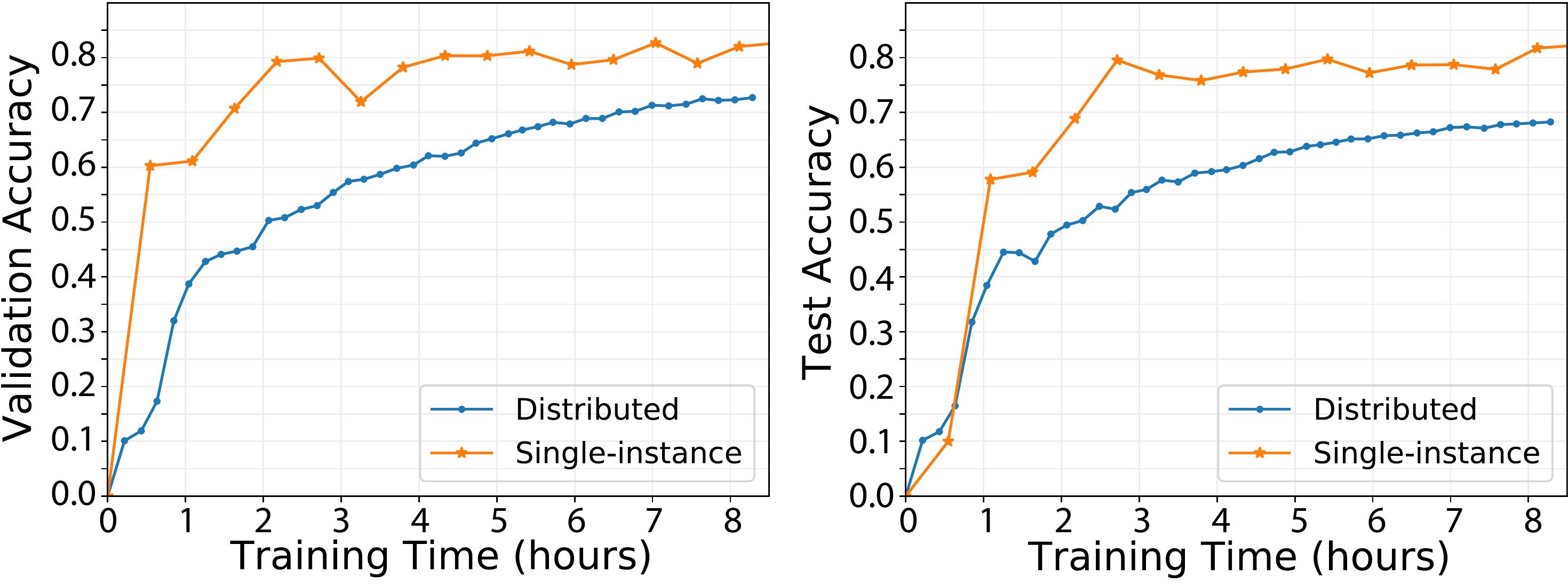}
	\caption{Validation (left) and Test (right) accuracy} \label{fig:baseline}
\end{figure}

To benchmark the performance of our distributed training approach against the best possible performance baseline, we run the CIFAR10 training job as a serial single-instance synchronous training. For the single-instance, we use the same configuration as the server computing instance. In practice, it is too slow to run a large training job on a single instance and, hence, the need for distributed training. 
Figure~\ref{fig:baseline} compares the performance of distributed training and single-instance training. The distributed curve is from P5C5T2 experiment with varying $\alpha$. The figure shows that test accuracy evolves similar to validation accuracy, which imparts confidence in our distributed training. We make three observations on our validation plot. First,  at the end of 8.4 hours, distributed training reaches an accuracy of 0.73 and the single-instance accuracy reaches 0.82. This result agrees with the prior work~\cite{zhang2015}, which shows that the distributed training accuracy is expected to be lower than the serial synchronous training accuracy at any given epoch. We did not perform a convergence analysis of our VC-ASGD scheme, similar to the one performed for EASGD~\cite{zhang2015}, and can not provide bounds on the accuracy gap between single-instance and distributed training schemes. However, past work~\cite{distbelief} has found that distributed training schemes can converge to acceptable accuracy levels. Tuning of hyperparameters, e.g. $\alpha$, is required to achieve convergence of  distributed training.
The second observation is that the accuracy gap between the two curves becomes smaller as the training time increases. This is promising because we can reduce the distributed training time by scaling the numbers of P, C and T (Section~\ref{sec:disttrain}) and by optimizing the subtask parameters such as the number of  epochs at a client. The third observation is that the distributed training curve is smoother and has less fluctuations than the single-instance curve. A smoother curve is desirable because it allows an easier quantification of the incremental gain in accuracy for an incremental increase in the training cost. 

\subsection{Impact of Eventual Consistency Database} \label{sec:dbeffect}
In our design, multiple parameter servers concurrently access and update a shared copy of server parameter values via Redis, a main-memory eventual consistency key-value store.  We store all the parameters of a model as a single value. We choose an eventual consistency database to improve scalability as described in Section~\ref{sec:concurrency}. To assess the impact of our choice, we compare the effect of storing the copy of server parameter values in Redis versus MySQL, a strong consistency database. We repeat the experiments described in Section~\ref{sec:disttrain} using MySQL. We store all the parameters of a single model as a LONGBLOB in a MySQL table. A LONGBLOB is a binary object that can hold byte arrays of size up to 4GB. Our results show that a parameter update operation takes 1.29 seconds in MySQL and 0.87 seconds in Redis. Hence, a strong consistency database like MySQL takes 1.5 times longer for each update transaction. For CIFAR10 training over 40 epochs, there are $\sim$2,000 update operations. Using MySQL adds an overhead of 14 minutes training time. For larger training jobs, the overhead can be in hours. For example, the total training data size of a benchmark problem like ImageNet~\cite{deng2009imagenet} is 800 times the total training data size of CIFAR10. In case of ImageNet, the number of update operations for 40 epochs will be $\sim$1,600,000, which adds an overhead of 187 hours. 

\subsection{Impact of Preemptible Instances}
\input{spotclient}

%% file: spotclient.tex
We use preemptible instances to lower the cost of training. Consider the training cost associated with the experiment P5C5T2. We run this experiment on a fleet consisting of 5 computing instances and a total of 40 vCPUs and 160 GB RAM. If we use standard computing instances, the fleet will cost us \$1.67 per hour. With preemptible instances, it will cost us \$0.50 per hour, which is a saving of 70\%. For the P5C5T2 experiment with an 8 hour run time, we spend \$4 with preemptible instances and \$13.4 with standard instances. 

To reduce the time clients spend in processing training subtasks, we can use horizontal or vertical scaling. With preemptible instances, the cost of horizontal and vertical scaling can differ because smaller instances may be discounted more than larger instances or vice versa. For example, 10 smaller instances with 4 vCPUs and 16GB RAM each could cost less than 5 larger instances with 8 vCPUs and 32GB RAM each. 

Here we provide a rough estimate of the impact of using preemptible instances on overall training time. For preemptible instances on AWS, frequency of interruption represents the rate at which AWS has reclaimed instances in the past month\footnote{https://aws.amazon.com/ec2/spot/instance-advisor/}. It can be \textless5\%, 5\textendash10\%, 10\textendash15\%, 15\textendash20\% or \textgreater20\%, and can vary by instance type. The average frequency of interruption across all geographical regions and instance types is \textless5\%. All the instances we use for training have a frequency of interruption \textless5\%. We did not see any terminations during the 8 hour training period for the P5C5T2 experiment. We choose client instances from several instance pools with similar computing resources, and, hence, the termination of an instance is generally independent of the termination of another instance. 

We use the binomial probability distribution to model the impact of instance termination. We model the usage of compute instances as independent Bernoulli trials where the probability of an instance getting terminated is $p$. Let $t_e$ be the average execution time of a training subtask. If the result of a subtask is not received within the timeout period $t_o$, we reschedule the subtask, and the total execution time will increase to $t_e + t_o$. We denote the total number of subtasks for a training job (number of epochs $\!\times\!$ number of subtasks per epoch) with $n_s$, the total number of client instances with $n_c$, and the number of simultaneous subtasks per client instance with $n_{tc}$. The total number of subtasks that can accrue a timeout is $n \!=\! \dfrac{n_s}{n_c \!\times\! n_{tc}}$, and the expected number of subtasks that will accrue a timeout is $np$. The expected training time with timeouts is $np(t_e\!+\!t_o) \!+\! n(1\!-\!p)t_e$ or $nt_e \!+\! npt_o$. The term $npt_o$ represents the expected increase in training time.

For the P5C5T2 experiment, $n_c \!=\! 5$, $n_{tc} \!=\! 2$, $n_s \!=\! 2000$ and the total training time is slightly more than 8 hr. The average execution time of a subtask is $t_e \!\leq\! 2.4$ min and the timeout is set $t_o \!=\! 5$ min. With $p \!=\! 0.05$, the expected increase in training time is 50 min, and with $p \!=\! 0.20$, it will increase to 200 min.

%% file: discussion.tex
\section{Summary and Conclusions}\label{sec:discussion}
Making distributed training of deep learning (DL) models faster and cheaper is an open problem in commercializing AI technologies. We proposed and demonstrated the suitability of a distributed DL platform based on the Volunteer Computing (VC)-like paradigm as a solution to the problem. Our design addressed the three main challenges raised within the VC-like environment: fault tolerance, heterogeneity of clients and network latency. We also proposed a novel asynchronous parameter update scheme, VC-ASGD, that is convergent within the VC-like paradigm. We tested the effectiveness of the proposed system in improving the training time and model accuracy for the CIFAR10 image classification problem. In particular, we evaluated the impact of distributed training system parameters -- number of clients, number of parameter servers, number of subtasks per client, and the VC-ASGD hyperparameter -- on training time and accuracy, and showed that we can reduce the training time by 50\%. We also discussed the impact of the database used for storing model parameters on the training time. Finally, by running our system on preemptible instances in a commercial cloud environment, we achieved a 70-90\% reduction in the training cost compared to using standard computing instances in the cloud. Our approach did not sacrifice training speed and model accuracy, and provided stronger data security than training on traditional VC systems.
 


%% file: eydle-paper-IEEE.bbl
\begin{thebibliography}{10}
\providecommand{\url}[1]{#1}
\csname url@samestyle\endcsname
\providecommand{\newblock}{\relax}
\providecommand{\bibinfo}[2]{#2}
\providecommand{\BIBentrySTDinterwordspacing}{\spaceskip=0pt\relax}
\providecommand{\BIBentryALTinterwordstretchfactor}{4}
\providecommand{\BIBentryALTinterwordspacing}{\spaceskip=\fontdimen2\font plus
\BIBentryALTinterwordstretchfactor\fontdimen3\font minus
  \fontdimen4\font\relax}
\providecommand{\BIBforeignlanguage}[2]{{%
\expandafter\ifx\csname l@#1\endcsname\relax
\typeout{** WARNING: IEEEtran.bst: No hyphenation pattern has been}%
\typeout{** loaded for the language `#1'. Using the pattern for}%
\typeout{** the default language instead.}%
\else
\language=\csname l@#1\endcsname
\fi
#2}}
\providecommand{\BIBdecl}{\relax}
\BIBdecl

\bibitem{horovod18}
A.~Sergeev and M.~D. Balso, ``Horovod: fast and easy distributed deep learning
  in tensorflow,'' \emph{CoRR}, vol. abs/1802.05799, 2018.

\bibitem{petuum}
E.~P. Xing, Q.~Ho, W.~Dai, J.~K. Kim, J.~Wei, S.~Lee, X.~Zheng, P.~Xie,
  A.~Kumar, and Y.~Yu, ``Petuum: A new platform for distributed machine
  learning on big data,'' \emph{IEEE Transactions on Big Data}, vol.~1, no.~2,
  pp. 49--67, 2015.

\bibitem{turingnlg}
\BIBentryALTinterwordspacing
(2020, Feb.) Turing-nlg: A 17-billion-parameter language model by microsoft.
  [Online]. Available:
  \url{https://www.microsoft.com/en-us/research/blog/turing-nlg-a-17-billion-parameter-language-model-by-microsoft/}
\BIBentrySTDinterwordspacing

\bibitem{googleswitch2021}
W.~Fedus, B.~Zoph, and N.~Shazeer, ``Switch transformers: Scaling to trillion
  parameter models with simple and efficient sparsity,'' \emph{arXiv preprint
  arXiv:2101.03961}, 2021.

\bibitem{fakenews}
R.~Zellers, A.~Holtzman, H.~Rashkin, Y.~Bisk, A.~Farhadi, F.~Roesner, and
  Y.~Choi, ``Defending against neural fake news,'' \emph{arXiv preprint
  arXiv:1905.12616}, 2019.

\bibitem{Pan2010TransferLearning}
S.~J. {Pan} and Q.~{Yang}, ``A survey on transfer learning,'' \emph{IEEE
  Transactions on Knowledge and Data Engineering}, vol.~22, no.~10, pp.
  1345--1359, 2010.

\bibitem{kondo2009cost}
D.~Kondo, B.~Javadi, P.~Malecot, F.~Cappello, and D.~P. Anderson,
  ``Cost-benefit analysis of cloud computing versus desktop grids,'' in
  \emph{2009 IEEE International Symposium on Parallel \& Distributed
  Processing}.\hskip 1em plus 0.5em minus 0.4em\relax IEEE, 2009, pp. 1--12.

\bibitem{distbelief}
J.~Dean, G.~S. Corrado, R.~Monga, K.~Chen, M.~Devin, Q.~V. Le, M.~Z. Mao,
  M.~Ranzato, A.~Senior, P.~Tucker, K.~Yang, and A.~Y. Ng, ``Large scale
  distributed deep networks,'' in \emph{Proceedings of the 25th International
  Conference on Neural Information Processing Systems - Volume 1}, ser.
  NIPS'12.\hskip 1em plus 0.5em minus 0.4em\relax Red Hook, NY, USA: Curran
  Associates Inc., 2012, pp. 1223--1231.

\bibitem{adam}
T.~Chilimbi, Y.~Suzue, J.~Apacible, and K.~Kalyanaraman, ``Project adam:
  Building an efficient and scalable deep learning training system,'' in
  \emph{11th $\{$USENIX$\}$ Symposium on Operating Systems Design and
  Implementation ($\{$OSDI$\}$ 14)}, 2014, pp. 571--582.

\bibitem{dai2019bigdl}
J.~J. Dai, Y.~Wang, X.~Qiu, D.~Ding, Y.~Zhang, Y.~Wang, X.~Jia, C.~L. Zhang,
  Y.~Wan, Z.~Li \emph{et~al.}, ``Bigdl: A distributed deep learning framework
  for big data,'' in \emph{Proceedings of the ACM Symposium on Cloud
  Computing}, 2019, pp. 50--60.

\bibitem{jsdoop}
J.~{\'A}. Morell, A.~Camero, and E.~Alba, ``Jsdoop and tensorflow. js:
  Volunteer distributed web browser-based neural network training,'' \emph{IEEE
  Access}, vol.~7, pp. 158\,671--158\,684, 2019.

\bibitem{kijsipongse2018hybrid}
E.~Kijsipongse, A.~Piyatumrong, U.~Suriya \emph{et~al.}, ``A hybrid gpu cluster
  and volunteer computing platform for scalable deep learning,'' \emph{The
  Journal of Supercomputing}, vol.~74, no.~7, pp. 3236--3263, 2018.

\bibitem{desell2017developing}
T.~Desell, ``Developing a volunteer computing project to evolve convolutional
  neural networks and their hyperparameters,'' in \emph{2017 IEEE 13th
  International Conference on e-Science (e-Science)}.\hskip 1em plus 0.5em
  minus 0.4em\relax IEEE, 2017, pp. 19--28.

\bibitem{ryabinin2020crowdsourced}
M.~Ryabinin and A.~Gusev, ``Towards crowdsourced training of large neural
  networks using decentralized mixture-of-experts,'' in \emph{Advances in
  Neural Information Processing Systems}, H.~Larochelle, M.~Ranzato,
  R.~Hadsell, M.~F. Balcan, and H.~Lin, Eds., vol.~33.\hskip 1em plus 0.5em
  minus 0.4em\relax Curran Associates, Inc., 2020, pp. 3659--3672.

\bibitem{foldingathome}
M.~I. Zimmerman, J.~R. Porter, M.~D. Ward, S.~Singh, N.~Vithani, A.~Meller,
  U.~L. Mallimadugula, C.~E. Kuhn, J.~H. Borowsky, R.~P. Wiewiora
  \emph{et~al.}, ``Citizen scientists create an exascale computer to combat
  covid-19,'' \emph{BioRxiv}, 2020.

\bibitem{boinc}
D.~P. Anderson, ``Boinc: A system for public-resource computing and storage,''
  in \emph{Fifth IEEE/ACM international workshop on grid computing}.\hskip 1em
  plus 0.5em minus 0.4em\relax IEEE, 2004, pp. 4--10.

\bibitem{bhattacharjee2017ibm}
B.~Bhattacharjee, S.~Boag, C.~Doshi, P.~Dube, B.~Herta, V.~Ishakian,
  K.~Jayaram, R.~Khalaf, A.~Krishna, Y.~B. Li \emph{et~al.}, ``Ibm deep
  learning service,'' \emph{IBM Journal of Research and Development}, vol.~61,
  no. 4/5, pp. 10:1--10:11, 2017.

\bibitem{bottou-curtis-nocedal2018_SGD}
L.~{Bottou}, F.~E. {Curtis}, and J.~{Nocedal}, ``Optimization methods for
  large-scale machine learning,'' \emph{Siam Reviews}, vol.~60, no.~2, pp.
  223--311, 2018.

\bibitem{LiParamServer_2014}
M.~Li, D.~G. Andersen, J.~W. Park, A.~J. Smola, A.~Ahmed, V.~Josifovski,
  J.~Long, E.~J. Shekita, and B.-Y. Su, ``Scaling distributed machine learning
  with the parameter server,'' in \emph{11th {USENIX} Symposium on Operating
  Systems Design and Implementation ({OSDI} 14)}.\hskip 1em plus 0.5em minus
  0.4em\relax Broomfield, CO: {USENIX} Association, Oct. 2014, pp. 583--598.

\bibitem{hogwild}
B.~Recht, C.~Re, S.~Wright, and F.~Niu, ``Hogwild!: A lock-free approach to
  parallelizing stochastic gradient descent,'' in \emph{Advances in Neural
  Information Processing Systems}, J.~Shawe-Taylor, R.~Zemel, P.~Bartlett,
  F.~Pereira, and K.~Q. Weinberger, Eds., vol.~24.\hskip 1em plus 0.5em minus
  0.4em\relax Curran Associates, Inc., 2011, pp. 693--701.

\bibitem{TensorflowOSDI2016}
M.~Abadi, P.~Barham, J.~Chen, Z.~Chen, A.~Davis, J.~Dean, M.~Devin,
  S.~Ghemawat, G.~Irving, M.~Isard, M.~Kudlur, J.~Levenberg, R.~Monga,
  S.~Moore, D.~G. Murray, B.~Steiner, P.~Tucker, V.~Vasudevan, P.~Warden,
  M.~Wicke, Y.~Yu, and X.~Zheng, ``Tensorflow: A system for large-scale machine
  learning,'' in \emph{Proceedings of the 12th USENIX Conference on Operating
  Systems Design and Implementation}.\hskip 1em plus 0.5em minus 0.4em\relax
  USA: USENIX Association, 2016, pp. 265--283.

\bibitem{HoStaleSynchronous2013}
Q.~Ho, J.~Cipar, H.~Cui, J.~K. Kim, S.~Lee, P.~B. Gibbons, G.~A. Gibson, G.~R.
  Ganger, and E.~P. Xing, ``More effective distributed ml via a stale
  synchronous parallel parameter server,'' \emph{Advances in neural information
  processing systems}, pp. 1223--1231, 2013.

\bibitem{Lian2015DelayedGradient}
X.~Lian, Y.~Huang, Y.~Li, and J.~Liu, ``Asynchronous parallel stochastic
  gradient for nonconvex optimization,'' in \emph{Advances in Neural
  Information Processing Systems}, C.~Cortes, N.~Lawrence, D.~Lee, M.~Sugiyama,
  and R.~Garnett, Eds., vol.~28.\hskip 1em plus 0.5em minus 0.4em\relax Curran
  Associates, Inc., 2015.

\bibitem{zhang2015}
S.~Zhang, A.~E. Choromanska, and Y.~LeCun, ``Deep learning with elastic
  averaging sgd,'' in \emph{Advances in Neural Information Processing Systems},
  C.~Cortes, N.~Lawrence, D.~Lee, M.~Sugiyama, and R.~Garnett, Eds.,
  vol.~28.\hskip 1em plus 0.5em minus 0.4em\relax Curran Associates, Inc.,
  2015, pp. 685--693.

\bibitem{zheng2017}
S.~Zheng, Q.~Meng, T.~Wang, W.~Chen, N.~Yu, Z.-M. Ma, and T.-Y. Liu,
  ``Asynchronous stochastic gradient descent with delay compensation,'' in
  \emph{International Conference on Machine Learning}.\hskip 1em plus 0.5em
  minus 0.4em\relax PMLR, 2017, pp. 4120--4129.

\bibitem{pmlr-v48-hardt16}
M.~Hardt, B.~Recht, and Y.~Singer, ``Train faster, generalize better: Stability
  of stochastic gradient descent,'' in \emph{Proceedings of The 33rd
  International Conference on Machine Learning}, ser. Proceedings of Machine
  Learning Research, M.~F. Balcan and K.~Q. Weinberger, Eds., vol.~48.\hskip
  1em plus 0.5em minus 0.4em\relax New York, USA: PMLR, 20--22 Jun 2016, pp.
  1225--1234.

\bibitem{Sermanet2014Cifar10}
P.~Sermanet, D.~Eigen, X.~Zhang, M.~Mathieu, R.~Fergus, and Y.~LeCun,
  ``Overfeat: Integrated recognition, localization and detection using
  convolutional networks,'' \emph{CoRR}, vol. 1312.6229, 2014.

\bibitem{He2016Resnet}
K.~He, X.~Zhang, S.~Ren, and J.~Sun, ``Deep residual learning for image
  recognition,'' in \emph{IEEE Conference on Computer Vision and Pattern
  Recognition (CVPR)}, 06 2016, pp. 770--778.

\bibitem{deng2009imagenet}
J.~Deng, W.~Dong, R.~Socher, L.-J. Li, K.~Li, and L.~Fei-Fei, ``Imagenet: A
  large-scale hierarchical image database,'' in \emph{2009 IEEE conference on
  computer vision and pattern recognition}.\hskip 1em plus 0.5em minus
  0.4em\relax Ieee, 2009, pp. 248--255.

\end{thebibliography}
